\newcommand\apj{{ApJ\,}}%
\newcommand\apjl{{ApJ\,}}%
\newcommand\apjs{{ApJS\,}}%
\newcommand\aap{{A\&A\,}}%
\newcommand\aaps{{A\&AS\,}}%
\newcommand\mnras{{MNRAS\,}}%
\newcommand\nat{{Nature\,}}%
\def\Dnu{$\Delta\nu$}
\def\numax{$\nu_{\rm max}$}

\def\kepler{\textit{Kepler}}
\def\Teff{$T_{\rm eff}$}
\documentclass[graybox, natbib, footinfo]{svmult}

\usepackage{mathptmx}       
\usepackage{helvet}         
\usepackage{courier}        
\usepackage{type1cm}        
\usepackage{makeidx}         
\usepackage{graphicx}        
\usepackage{multicol}        
\usepackage[bottom]{footmisc}

\usepackage{lscape}
\usepackage{amssymb}
\usepackage{url}
\usepackage{journals}

\makeindex             

\begin{document}
\title*{Asteroseismology of red giants as a tool for studying stellar populations: first steps}
\author{Andrea Miglio}
\institute{A. Miglio \at School of Physics and Astronomy, University of Birmingham, United Kingdom\\
Institut d'Astrophysique et de G\'eophysique de l'Universit\'e de Li\`ege, Belgium\\
 \email{miglioa@bison.ph.bham.ac.uk}
}
%
%
\maketitle

\abstract{
The detection of solar-like oscillations in G and K giants with the CoRoT and \kepler\ space-based satellites allows robust constraints to be set on the mass and radius of such stars. The availability of these constraints for thousands of giants sampling different regions of the Galaxy promises to enrich our understanding on the Milky Way's constituents. In this contribution we briefly recall which are the relevant constraints that red-giants seismology can currently provide to the study of stellar populations. We then present, for a few nearby stars, the comparison between radius and mass determined using seismic scaling relations and those obtained by other methods.}
\section{Introduction}
\label{sec:intro}
Since the data from the first CoRoT observational runs were analysed, and solar-like oscillations were detected in thousands of red giant stars \citep{DeRidder09,Hekker09, Mosser2010, Kallinger2010}, it has become clear that the newly available observational constraints will allow novel approaches in the study of so far poorly constrained galactic stellar populations \citep{Miglio2009}. 

While CoRoT continues to monitor giants in different regions of the Milky Way, \kepler\ is contributing significantly to the characterisation not only of red-giant populations (see De Ridder, this volume for a review) but it has also opened the way for ``ensemble seismology'' of solar-like stars. The detection of solar-like oscillations in about 500 F and G dwarfs allowed \citet{Chaplin2011} to perform a first quantitative comparison between the distributions of observed masses and radii of these stars with predictions from models of synthetic populations in the Galaxy.

We outline in Sec. \ref{sec_stpop} the innovative aspects of seismic constraints, highlighting the importance of being able to determine the mass of giant stars, while we will discuss in detail the implications of the radius (hence distance) estimates in a future paper. 
In Sec. \ref{sec:scale} we first review how mass and radius of giants are estimated using the average seismic parameters  \Dnu\ (average large frequency separation) and  \numax\ (frequency corresponding to the maximum observed oscillation power), and then present, for a few nearby stars, the comparison between radius and mass determined using seismic and non-seismic observational constraints.

\section{New constraints on stellar populations}
\label{sec_stpop}
Once they reach the red-giant phase of their evolution, stars of significantly different age end up sharing similar photospheric properties. As a consequence, field giants belonging to the composite galactic-disk population were so far considered poor tracers of age. However, the possibility of determining with asteroseismology the masses of thousands of these objects has unexpectedly reversed this picture. 

\begin{figure*}
\centering
   \includegraphics[width=.7\hsize]{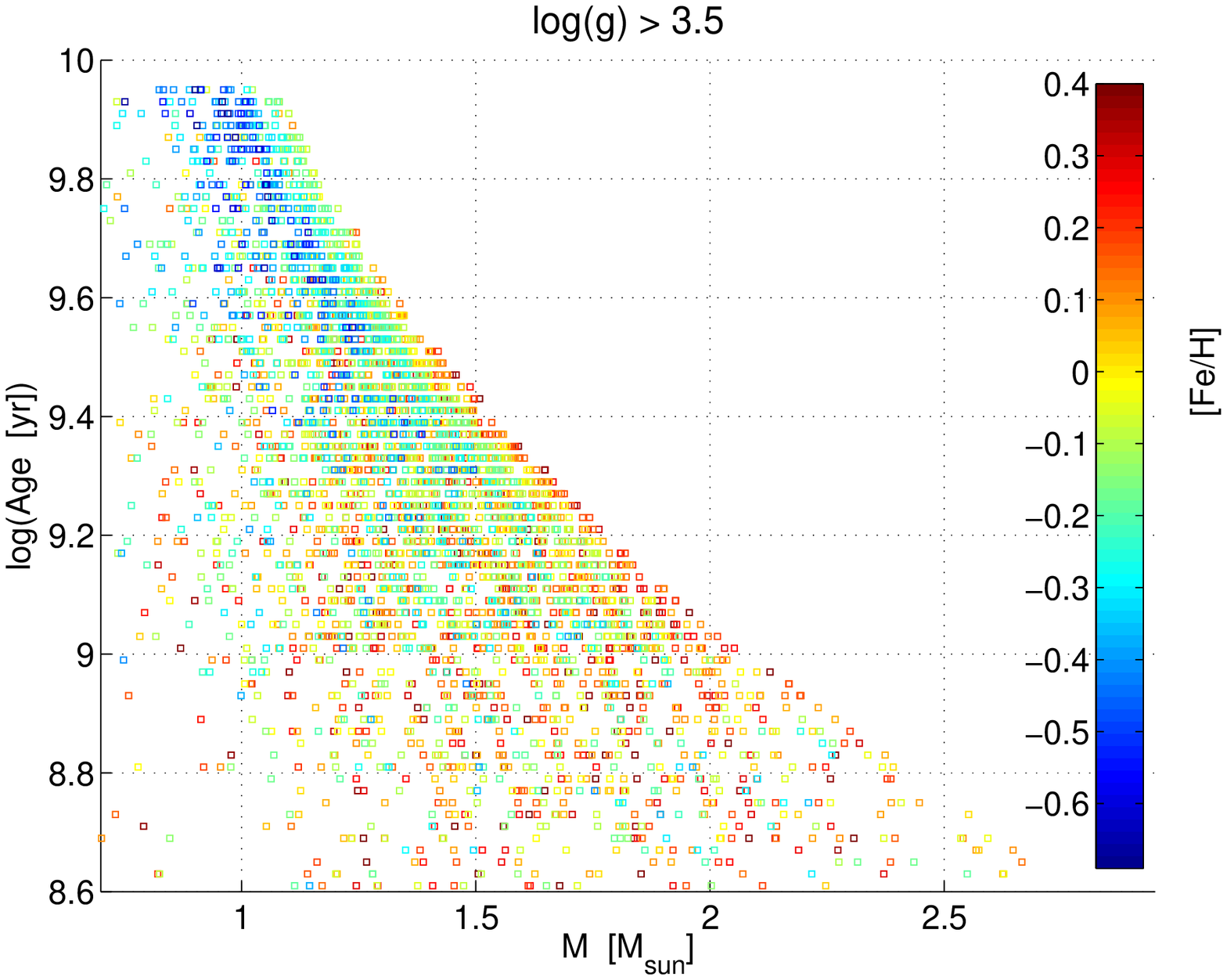}
      \includegraphics[width=.7\hsize]{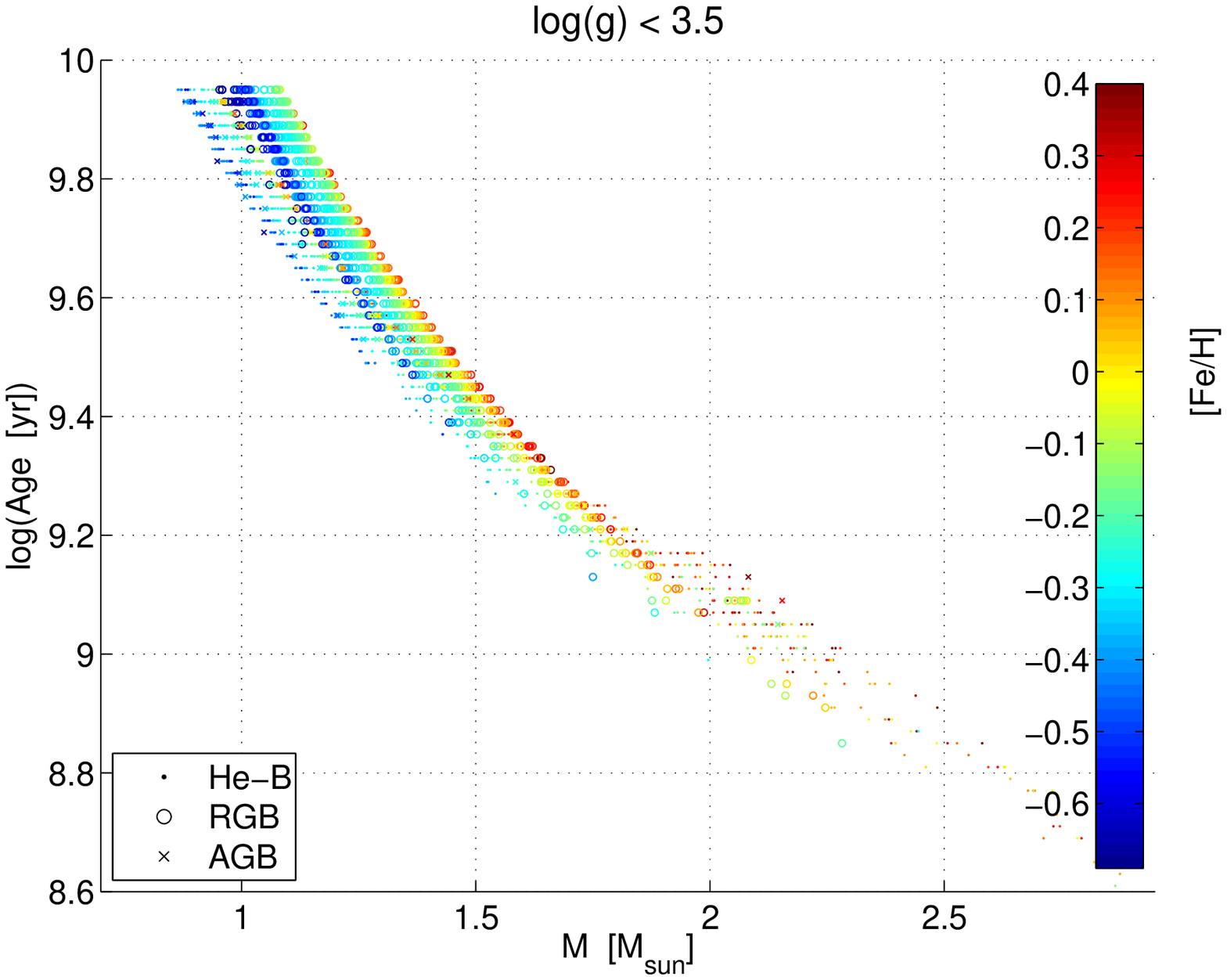}
      \caption{Age-mass-metallicity relation for  main-sequence stars ({\it upper panel})  and red giants ({\it lower panel}) in a synthetic population representative of thin-disk stars observed by CoRoT in the LRc01 field. The evolutionary state of giants is marked with a different symbol: dots (stars in the core-Helium-burning phase), crosses (Asymptotic-Giant-Branch stars), and open circles (stars on the Red Giant Branch). The fraction of AGB stars in the population of giants shown here is $\sim4\%$.}
         \label{fig:age}
\end{figure*}

As is well known, the age of  RGB and red-clump (RC) stars is largely determined by their main-sequence lifetime and hence, to a first approximation, by their mass and metallicity. The age-mass relation of giant stars predicted by stellar models is illustrated in Fig. \ref{fig:age}, where it is compared with that of stars on the main sequence. For the purposes of this comparison, a crude criterion based on the surface gravity $g$ was used to separate giants ($\log{g} < 3.5$) from main-sequence stars ($\log{g} > 3.5$). The synthetic population shown in the figure was computed with the code \mbox{TRILEGAL} (\citealt{Girardi05}; Girardi et al., this volume), and is representative of thin-disk stars monitored by CoRoT in the LRc01 field.
The tight age-mass relation shown in the lower panel of Fig.  \ref{fig:age} clearly shows that adding the mass among the observational constraints enables us to use giants as potentially very precise age indicators.
 
From a closer inspection of Fig. \ref{fig:age}, it is worth noticing  that for stars with $M \lesssim 1.5$ $M_\odot$ the age-mass relation bifurcates due to the significant  mass loss ($\sim 0.1-0.2$ $M_\odot$) experienced by low-mass stars near the tip of the RGB\footnote{In the models used in Fig. \ref{fig:age}, RGB mass loss is implemented adopting the \citet{Reimers1975a} prescription (see \citealt{Girardi2000} for more details).}.  Consequently, RC stars are younger than stars on the  RGB with the same actual mass (and metallicitiy). We can, however, remove this degeneracy in the age-mass relation thanks to additional seismic constraints. It is indeed excellent news in this context that the detailed properties of dipolar oscillation modes allow us to clearly distinguish RGB from RC stars (\citealt{Montalban2010, Bedding2011, Mosser2011}; Montalb\'an et al, this volume). When applied to the characterisation of stellar populations this result can potentially lead to age estimates independent of the uncertain RGB mass-loss rates.

As a word of caution we should recall, however, that these age estimates are inherently model dependent, being affected by uncertainties in predicting, e.g., main-sequence lifetimes. On the other hand, the potential of asteroseismology goes well beyond the determination of global stellar parameters using scaling relations. As frequencies of individual pulsation modes become available,  detailed comparisons between observed and theoretical oscillation spectra promise to improve both the precision of age estimates (see e.g. \citealt{DiMauro2011}), along with their accuracy, by providing stringent constraints on models of the internal structure of both main-sequence and in giant stars. 

As a relevant additional constraint that seismology could potentially provide to the study of stellar populations, we recall that investigations are currently underway to assess under which conditions a reliable indication of the envelope-helium abundance can be derived from the seismic signature of helium ionisation detected in CoRoT and \kepler\ giants (see \citealt{Miglio2010}; Montalb\'an et al., this volume).

Finally, as discussed during this meeting, it is worth mentioning that  Eq. \ref{eq:numax} below provides a potentially very accurate way of determining the surface gravities of stars, which could be then used as an input to refine  spectroscopic analyses (see e.g. \citealt{Morel2011}) and, eventually, to test model atmospheres of giant stars (Plez, this meeting).


\section{Scaling relations}
\label{sec:scale}
Radii and masses of solar-like oscillating stars can be estimated from the average seismic parameters that characterise their oscillation spectra: the so-called average large frequency separation (\Dnu), and the frequency corresponding to the maximum observed oscillation power (\numax).

The large frequency separation is predicted by theory to scale as the square root of the mean density of the star \citep[see e.g.][]{Vandakurov67, Tassoul80}:
\begin{equation}
\Delta\nu\simeq\sqrt{\frac{M/M_\odot}{(R/R_\odot)^3}}\Delta\nu_{\odot}{\rm,}
\label{eq:dnu}
\end{equation}
where $\Delta\nu_\odot=135$ $\mu$Hz. The frequency of maximum power is expected to be proportional to the acoustic cutoff frequency \citep{Brown1991, Kjeldsen1995, Mosser2010, Belkacem2011}, and therefore:
\begin{equation}
\nu_{\rm max}\simeq\frac{M/M_\odot}{(R/R_\odot)^2\sqrt{T_{\rm eff}/T_{\rm eff,\odot}}}\nu_{\rm max,\odot}\,{\rm,}
\label{eq:numax}
\end{equation}
where $\nu_{\rm max,\odot}=3100$ $\mu$Hz and $T_{\rm eff,\odot}=5777$ K.

Depending on the observational constraints available,  we may derive mass estimates from Equations \ref{eq:dnu} and \ref{eq:numax} alone, or via their combination with other available information from non-seismic observations.  When no information on distance/luminosity is available, which is the case for the vast majority of field stars observed by CoRoT and \kepler, Eq. \ref{eq:dnu} and \ref{eq:numax} may be solved to derive $M$ and $R$ \citep[see e.g.][]{Kallinger2010, Mosser2010}:
\begin{eqnarray}
\frac{M}{M_\odot} &\simeq& \left(\frac{\nu_{\rm max}}{\nu_{\rm max, \odot}}\right)^{3}\left(\frac{\Delta\nu}{\Delta\nu_{\odot}}\right)^{-4}\left(\frac{T_{\rm eff}}{T_{\rm eff, \odot}}\right)^{3/2} \label{eq:scalM}      \\
\frac{R}{R_\odot} &\simeq& \left(\frac{\nu_{\rm max}}{\nu_{\rm max, \odot}}\right)\left(\frac{\Delta\nu}{\Delta\nu_{\odot}}\right)^{-2}\left(\frac{T_{\rm eff}}{T_{\rm eff, \odot}}\right)^{1/2}{\rm.}\label{eq:scalR}
\end{eqnarray}

However, when additional constraints on the distance/luminosity of stars are available, $M$ can also be estimated also from Eq. \ref{eq:dnu} or \ref{eq:numax} alone:

\begin{eqnarray}
\frac{M}{M_\odot} &\simeq& \left(\frac{\Delta\nu}{\Delta\nu_{\odot}}\right)^{2}\left(\frac{L}{L_\odot}\right)^{3/2}   \left(\frac{T_{\rm eff}}{T_{\rm eff, \odot}}\right)^{-6} \label{eq:scalM2}      \\
\frac{M}{M_\odot} &\simeq& \left(\frac{\nu_{\rm max}}{\nu_{\rm max, \odot}}\right)\left(\frac{L}{L_{\odot}}\right)\left(\frac{T_{\rm eff}}{T_{\rm eff, \odot}}\right)^{-7/2}  \label{eq:scalM3}     
\end{eqnarray}

These scaling relations have been widely adopted to estimate masses and radii of red giants \citep[see e.g.][]{Stello2008,Kallinger2010, Mosser2010}, but they are based on simplifying assumptions which must be checked against independent fundamental measurements. Recent advances have been made on providing a theoretical basis for the relation between the acoustic cut-off frequency and \numax\  \citep{Belkacem2011}, and preliminary investigations with stellar models \citep{Stello2009a} indicate that the scaling relations hold to within $\sim3$\% on the main sequence and RGB (see also the Supporting Online Material in \citealt{Chaplin2011}).

\subsection{Empirical tests of the \numax\ and \Dnu\ scaling relations}
To assess the accuracy of the scaling relations,  ongoing studies based on models of stars in different evolutionary phases, and covering a wide range of  parameters \cite[see e.g.][]{White2011,Miglio2011}, must be complemented by calibration of the \numax\ and \Dnu\ relations with independent determinations of masses and radii. As a very first step in this process, we present here a simple comparison between radii and masses determined via seismic constraints with those obtained by other methods (combination of parallax, bolometric flux, effective temperature, angular radius, mass derived from the orbital solution of binary systems).

\begin{figure*}
\centering
   \includegraphics[width=\hsize]{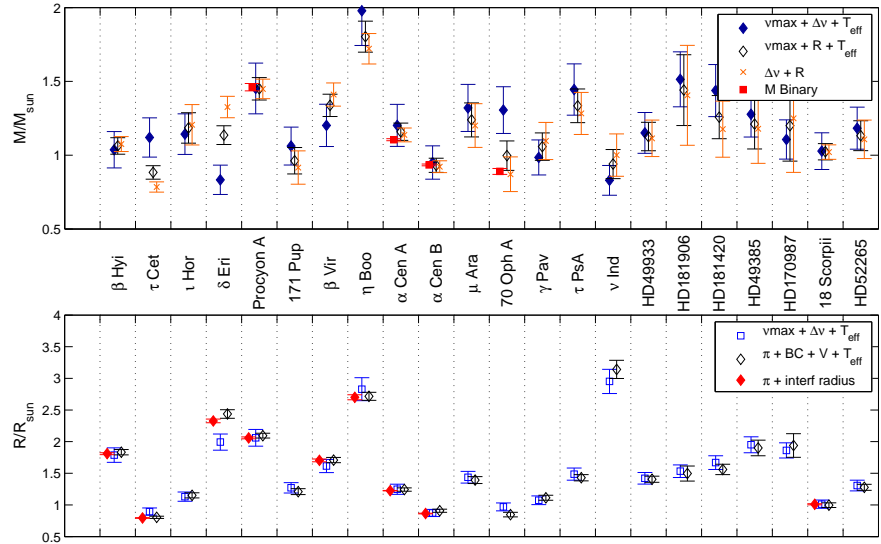}
      \caption{Comparison between masses ({\it upper panel}) and radii {\it upper panel} determined by different combinations of the observational constraints available.}
         \label{fig:alldwarfs}
\end{figure*}

\begin{figure*}
\centering
   \includegraphics[width=\hsize]{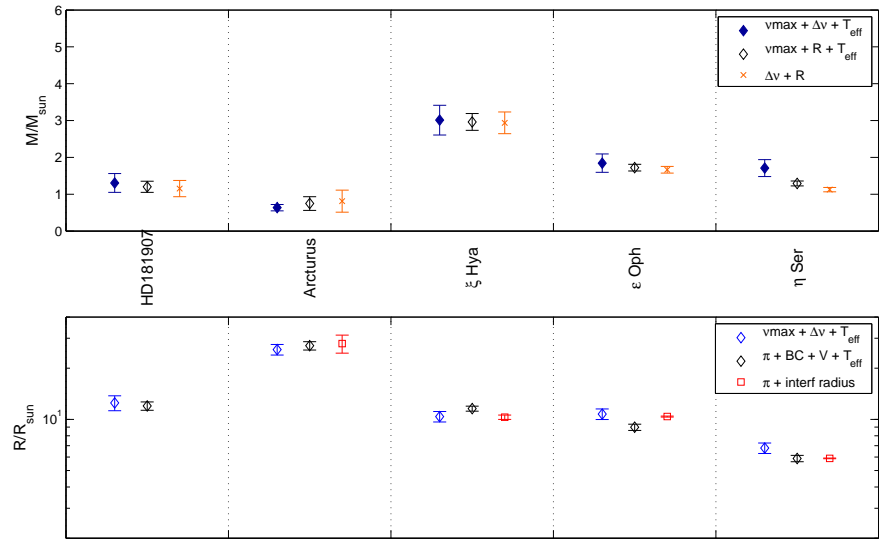}
      \caption{Same as Fig. \ref{fig:alldwarfs}, but considering giants with published seismic analysis.}
         \label{fig:allgiants}
\end{figure*}



We include in this comparison nearby stars with available seismic constraints, along the lines of the work presented by \citet{Bruntt2010}. 
We consider a total of 27 stars with published values of both \numax\ and \Dnu.  The quality of the seismic data available for the stars in this sample is highly heterogeneous, ranging from nearly 6-months long space-based photometric observations with the CoRoT satellite, to few days' single-site radial-velocity monitoring. The methods used to estimate \numax\ and \Dnu\ are also not uniform. We therefore decided to adopt a 2\% and 5\% uncertainty in \Dnu\ and \numax, respectively, as also suggested in \citet{Bruntt2010}.

Asteroseismic, spectroscopic, interferometric, and photometric constraints were either taken from the \citet{Bruntt2010} compilation (to which we refer for the original references), or collected from the papers by \citet{Ballot2011, Barban2009, Bazot2011,Bruntt2009, Carrier2006,Carrier10, Deheuvels2010, Eggenberger2008,  Gillon2006,Kallinger2010, Mathur2010, Mazumdar2009, Merand2010,Mosser2008,Mosser2009,Mosser2010}; and  \citet{Quirion2010}. Parallaxes are taken from \citet{VanLeeuwen2007} and bolometric corrections from \citet{Flower1996}. 
When available, we used \Teff\ determined from the bolometric fluxes and interferometric angular radii, in which case we considered the value quoted in \citet{Bruntt2010}. Otherwise, we adopted  spectroscopic \Teff\ with uncertainties of 100~K, unless the uncertainty was larger in the original reference.  As in \citet{Bruntt2009} we excluded from the sample HD175726 since its estimated large separation shows an unexplained large modulation with frequency \citep[see][]{Mosser2009}.

We then determined radii using Eq. \ref{eq:scalR} and masses via Eq. \ref{eq:scalM} or by including constraints on the luminosity, using Eqs. \ref{eq:scalM2} and \ref{eq:scalM3}. The comparisons between different (not always independent) determinations of radius and mass are presented in Figures \ref{fig:alldwarfs} and \ref{fig:allgiants}.
\begin{figure*}
\centering
   \includegraphics[width=.85\hsize]{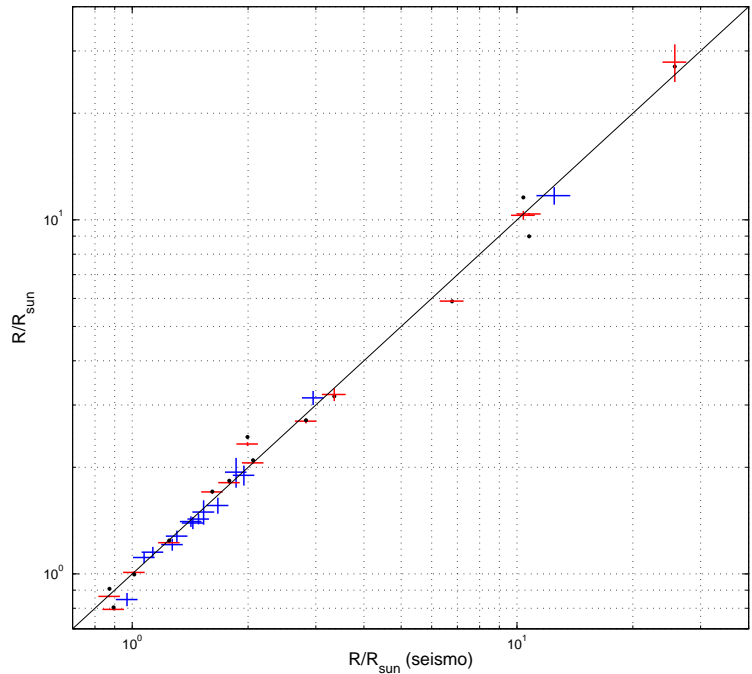}
      \caption{Radii determined using seismic constraints (Eq. \ref{eq:scalR}) vs. radii determined from parallax and angular interferometric radius (red), and from apparent magnitude, parallax, BC and $T_{\rm eff}$ (blue). For targets where interferometric measurements are available we also report the comparison with radii determined from apparent magnitude+ parallax+ BC + $T_{\rm eff}$ (black dots).}
         \label{fig:allr}
\end{figure*}

\begin{figure*}
\centering
   \includegraphics[width=\hsize]{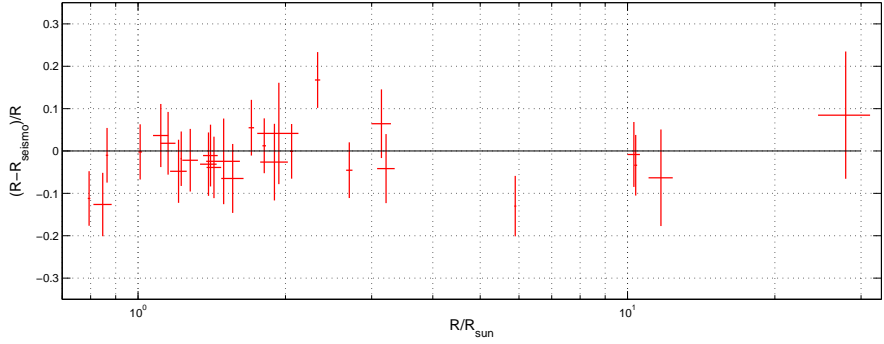}
      \caption{Percentage difference between radii determined using seismic constraints (Eq. \ref{eq:scalR}) and those derived from the parallax and interferometric angular radius (if available) or the the estimated bolometric luminosity and \Teff.}
         \label{fig:allr2}
\end{figure*}
The targets considered span a large domain in radius: from sub-solar radii  ($\tau$ Cet, 70 Oph A, and $\alpha$ Cen B) to the $\sim30$ $R_\odot$ of the metal-poor giant Arcturus. The overall agreement found between values determined via Eq. \ref{eq:scalR} and using classical constraints is remarkable (see Fig. \ref{fig:allr} and \ref{fig:allr2}), and the two determinations agree within 1-$\sigma$ ($\sim7\%$) in most cases (see Fig. \ref{fig:allr2}). Weighting the differences according to their errors, we find a mean difference ($R_{\rm seismo}-R$) and standard deviation of -1.5\% and 6\%, respectively. A significantly larger number of stars (especially giants) should be included to investigate possible trends with stellar properties. This comparison is indeed encouraging and adds strong support to the use of solar-like oscillators as distance indicators, also when compared to results obtained using accurate determinations of radii in eclipsing binaries, as presented in the recent review by \citet{Torres2009} (see e.g. their Fig. 1).

\begin{figure*}
\centering
   \includegraphics[width=\hsize]{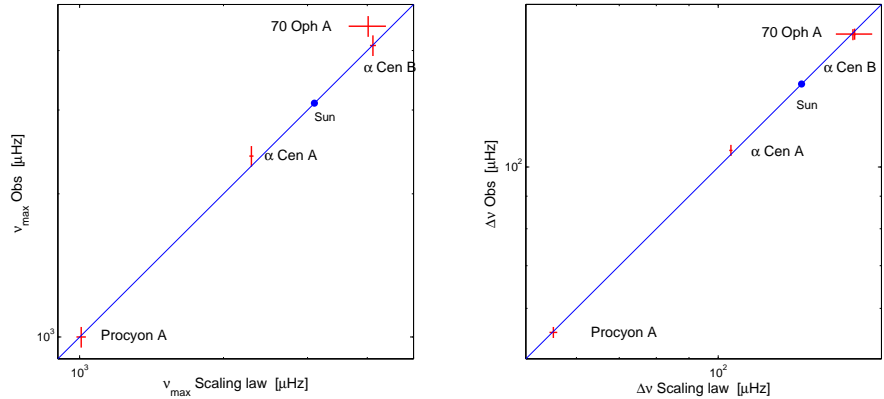}
      \caption{{\it Left panel}: Comparison between \numax\ observed  vs. \numax\ predicted using independent measurements of mass and radius.  {\it Right panel}: as left panel, but for \Dnu.}
         \label{fig:numaxdnu}
\end{figure*}

The expected uncertainty in the mass determined using Eq. \ref{eq:scalM} and the available data is $\sim10-15\%$. Besides noting a good agreement when comparing different (but correlated) expressions to estimate the mass (see Fig. \ref{fig:alldwarfs} and \ref{fig:allgiants}), only for few visual binary systems ($\alpha$ Cen A and B, Procyon A, and 70 Oph A) could we test the mass determined using \numax\ and/or \Dnu\ with the independent estimate based on the orbital solution. In these cases we find a 1-$\sigma$ agreement, except for 70 Oph A which has an observed \numax\ larger than expected (still within 2 $\sigma$  of the predicted value).  This is clear from the direct comparison between \numax\ and \Dnu\ observed and predicted by scaling relations is shown in Fig. \ref{fig:numaxdnu}.

The quality of seismic constraints obtained from space-based data exceeds that available for most of the stars considered in this comparison. Consequently, while radii and masses can be estimated with greater precision, this demands more stringent tests of the accuracy of the scaling relations.
In this respect nearby  \kepler\ and CoRoT targets, and in particular high-duty cycle ground-based observations (e.g. with SONG, see \citealt{Grundahl2009}) will play a crucial role in testing \numax\ and \Dnu\ in well constrained systems.
Moreover, the detection with \kepler\ of solar-like oscillations in red giants members of open clusters provides additional means for testing the accuracy of scaling relations, particularly when largely model-independent constraints are available for cluster members (see the encouraging results reported in \citealt{Stello2010, Basu2011, Miglio2011}).
Eclipsing binaries with solar-like pulsating components observed by \kepler\ and CoRoT are also promising and privileged targets for this purpose  \citep[see e.g.][]{Hekker2010}.

\section{Summary \& outlook}
Thanks to the interpretation of solar-like oscillation spectra detected by CoRoT and \kepler, we can now determine the mass and radius of thousands of stars belonging to the composite population of the Milky Way's disk. These truly innovative constraints will allow precise age estimates for giants, and will inform studies of galactic formation and evolution with observational constraints which were not available prior to asteroseismology. To fully exploit the potential of these observations, however, it will be crucial to combine them with spectroscopic constraints, which should become available in the near future thanks to large spectroscopic surveys such as SDSS-APOGEE and the GAIA-ESO spectroscopic survey. Further efforts should also be devoted to assess the validity of the \numax\ and \Dnu\ scaling relations, both in terms of their theoretical foundation,  and through calibration with independent measurements of radius and mass.

In the future, the pioneering observations of CoRoT and \kepler\ could be extended  to significantly wider areas of the sky by the candidate ESA mission PLATO\footnote{\texttt{http://sci.esa.int/plato}}, 
providing observational constraints that will be complementary to the accurate distance and proper motions measured by GAIA\footnote{\texttt{http://gaia.esa.int/}} in the coming years.  
\begin{acknowledgement}
The author acknowledges FNRS for financial support, M. Barbieri,  L. Girardi, J. Montalb\'an, T. Morel, B. Mosser, and A. Noels for enlightening discussions about seismology and stellar populations.  Additional thanks are due to M. Barbieri and L. Girardi for their kind help with the code TRILEGAL, and to W.J. Chaplin for reading the manuscript.
\end{acknowledgement}
%


\end{document}